\documentclass[preprint,eqsecnum,nofootinbib,natbib,aps]{revtex4}
\usepackage{graphicx}

\begin{document}
\title{Inflationary Solutions in  Nonminimally
Coupled Scalar Field Theory}
\author{Seoktae Koh}
\email{kohst@ihanyang.ac.kr}
\affiliation{Department of Science Education, Ewha Womans University, Seoul
120-750, Korea}
\author{Sang Pyo Kim}
\email{sangkim@kunsan.ac.kr}
\affiliation{Department of Physics, Kunsan National University, Kunsan
573-701, Korea and\\
Asia Pacific Center for Theoretical Physics, Pohang 790-784, Korea}
\author{Doo Jong Song}
\email{djsong@kao.re.kr}
\affiliation{ Korea Astronomy Observatory, Daejeon 305-348, Korea}

\begin{abstract}
We study analytically and numerically the inflationary solutions
for various type scalar potentials in the nonminimally coupled
scalar field theory. The Hamilton-Jacobi equation is used to deal
with nonlinear evolutions of inhomogeneous spacetimes and the
long-wavelength approximation is employed. The constraints that
lead to a sufficient inflation are found for the nonminimal
coupling constant and initial conditions of the scalar field for
inflation potentials. In particular, we numerically find an
inflationary solution in the new inflation model of a nonminimal
scalar field.
\end{abstract}


\maketitle

\section{Introduction}

The origin of the large scale structure could be well explained in
the inflationary scenario which predicts a scale invariant
spectrum, Gaussian statistics, and the curvature perturbation.
However, a recent attention on the non-Gaussianity of the
temperature anisotropy \cite{maldacena03} has motivated the
investigation of the gravitational perturbations during an
inflation period beyond the linear order theory.  The
Hamilton-Jacobi theory, for instance, has been used to study
nonlinear evolutions of inhomogeneous spacetimes in Einstein
gravity \cite{salopek90} and generalized gravity
\cite{soda95,koh05}. Even though it is difficult to get exact
solutions of the Hamilton-Jacobi equation, the long-wavelength
approximation is useful in dealing with superhorizon size
perturbations. When the scale of perturbations is larger than the
horizon size, spatial gradient terms can be neglected compared to
their temporal variations. In this sense, the lowest order
Hamilton-Jacobi and evolution equations for fields look like
homogeneous equations of motion in the long-wavelength
approximation.

Inflation potentials that dominate the energy density during the
inflation period, are a crucial ingredient to discriminate among
different inflation models. In Ref. \cite{lidsey97}, the inflation
potentials were reconstructed from observational data. The
inflation models can be classified according to the shape of the
potential and the initial condition of the scalar field into three
types: a large field, small field and hybrid field model
\cite{dodelson97}. The chaotic inflation \cite{linde83}, which
belongs to the large field model, has a positive curvature at the
minimum of the potential and needs the initial condition $\phi_0 >
m_{pl}$ to result in a sufficient inflation. Contrary to the
chaotic inflation, the new inflation \cite{linde82} (a small field
model) has a negative curvature at the false vacuum and requires
$\phi_0 \ll m_{pl}$. The scalar field in the new inflation starts
from the false vacuum initially and then rolls down to the true
vacuum. And the hybrid inflation \cite{linde91, copeland94} has a
positive curvature at the local minimum of the potential but has a
non-zero energy which is different from the chaotic inflation
model. One of the interesting features of the hybrid inflation is
prediction of a blue power spectrum ($n>1$, $n$ being a spectral
index) when perturbation modes leave the horizon.

The Brans-Dicke-like theories are naturally derived from the
fundamental physics theory such as string or M-theory. They are
also widely used to investigate the dark energy problem that is
believed to be responsible for the present accelerating Universe.
When the scalar fields are coupled to the spacetime curvature $R$
through $\xi R\phi^2/2$, the theory is renormalizable.  The
inflationary solutions were investigated in this nonminimally
coupled scalar field theory and constraints on the nonminimal
coupling constant $\xi$ were found \cite{futamase89, fakir90,
tsujikawa00}. But many of models considered there are the chaotic
inflation. The coupling term might prevent inflation from
occurring in the new inflation model \cite{abbott81} because it
behaves like a mass term in the scalar potential and destroys the
flatness of the potential.

In this paper we try to find inflationary solutions in the
nonminimally coupled scalar field theory for various potentials
and use the Hamilton-Jacobi theory to deal with nonlinear
evolutions of inhomogeneous spacetimes. Although the inflationary
solutions in the new inflation model are believed to be
impossible, we try to get the inflationary solutions by numerical
calculations and the slow-roll approximation in the new inflation.
We also find analytically or numerically the inflationary
solutions for chaotic and hybrid inflation models and put
constraints on the nonminimal coupling constant.

This paper is organized as follows. In. Sec. \ref{sec_hj}, we
derive the Hamilton-Jacobi and evolution equations for the
gravitation and scalar fields when the scalar field is
nonminimally coupled to the gravity. And then we solve the
Hamilton-Jacobi equation approximately using slow-roll conditions
during the inflation period in Sec. \ref{sec_slow}. We consider
the inflation potentials in chaotic, new, and hybrid inflation
models. To compare with approximate analytical solutions, we
perform numerical calculations in Sec. \ref{sec_num} and finally
we summarize our results in Sec. \ref{sec_summary}.

\section{Hamilton-Jacobi Equation in Nonminimally Coupled Scalar
Field Theory \label{sec_hj}}

The action for a nonminimally coupled scalar field takes the form
\begin{eqnarray}
\mathcal{I}=\int d^4 x\sqrt{-g}\left[(1-8\pi G\xi \phi^2)\frac{R}{16\pi
G} -\frac{1}{2}g^{\mu\nu}\partial_{\mu}\phi \partial_{\nu}\phi
-V(\phi)\right], \label{action}
\end{eqnarray}
where $\xi$ is a nonminimal coupling constant. The action
(\ref{action}) may be interpreted to have an effective
gravitational constant $G_{eff}$ depending on the scalar field as
\cite{futamase89}
\begin{eqnarray}
G_{eff} = \frac{G}{1-\phi^2/\phi_c^2}, \label{eff_grav}
\end{eqnarray}
where
\begin{equation}
\phi_c^2 = \frac{1}{ 8 \pi G \xi} \equiv \frac{m_{pl}^2}{8\pi
\xi}.
\end{equation}
We will require $G_{eff}>0$ to relate to our present Universe, and
thus we restrict $\phi$ to the region $\phi < \phi_c$ for $\xi
>0$. In the Hamilton-Jacobi formalism, the spacetime is written in
the ADM metric
\begin{eqnarray}
ds^2 = - N^2 dt^2 + \gamma_{ij}dx^i dx^j,
\end{eqnarray}
where $N$ and $\gamma_{ij}$ are a lapse function
and a 3-spatial metric  and we set a shift vector $N^i =0$.

The Hamilton-Jacobi theory proves useful for solving the
gravitational and scalar field equations which can be obtained
from the variation of the action (\ref{action}). By introducing a
generating functional $S$ that is a function of $\gamma_{ij},
\phi$ and integration constants, we can get the Hamilton-Jacobi
and momentum constraint equations \cite{salopek90,koh05}. The
generating functional, $S(\phi,\gamma_{ij})$, can be expanded in a
series in the order of spatial gradient terms
\begin{eqnarray}
S = S^{(0)} +S^{(2)} +S^{(4)}+\cdots.
\end{eqnarray}
The lowest order Hamilton-Jacobi equation is sufficient to deal
with the nonlinear evolution of the gravitational fields whose
scales are larger than the horizon size. We will assume an ansatz
for the lowest order generating functional of the form
\cite{koh05}
\begin{eqnarray}
S^{(0)}(\phi,\gamma_{ij}) = -\frac{1}{4\pi G} \int d^3 x
\gamma^{1/2}(1-8\pi G\xi\phi^2)^{3/2}W(\phi,\gamma_{ij}),
\end{eqnarray}
so that $S^{(0)}$ satisfies automatically the momentum constraint
equation. For Einstein gravity, $W$ can be interpreted as a
locally defined Hubble parameter. Then the Hamilton-Jacobi and
evolution equations for $\gamma_{ij}$ and $\phi$ are given by
\begin{eqnarray}
& &W^2 - \frac{4\pi G(1-8\pi G\xi \phi^2(1-6\xi))}{3(1-8\pi G \xi
\phi^2)^3} \left(\frac{1}{N}\dot{\phi}\right)^2 -\frac{8\pi
G}{3(1-8\pi G\xi \phi^2)^2}
V =0, \label{hj} \\
& &\frac{1}{N}\dot{\alpha} \equiv H = \sqrt{1-8\pi G\xi\phi^2}
\left[W + \frac{8\pi G\xi\phi}{(1-8\pi G\xi\phi^2)^{3/2}}\frac{1}{N}
\dot{\phi}\right],
\label{hubble} \\
& &\frac{1}{N}\dot{\phi} = -\frac{(1-8\pi G\xi\phi^2)^{5/2}} {4\pi
G(1-8\pi G\xi\phi^2(1-6\xi))}W_{\phi}, \label{field_eq}
\end{eqnarray}
where the metric is factored into a conformal part and a
unimodular metric
\begin{eqnarray}
\gamma_{ij} (t, {\bf x}) = e^{2\alpha(t,{\bf x})} h_{ij}({\bf x}),
\quad \det(h_{ij}) = 1.
\end{eqnarray}
There appears another singular point $\phi=\phi_s$ from Eq.
(\ref{field_eq}) in the case $0<\xi<1/6$, where
\begin{eqnarray}
\phi_{s} = \frac{m_{pl}}{\sqrt{8\pi \xi(1-6\xi)}} =
\frac{\phi_c}{\sqrt{1-6\xi}}.
\end{eqnarray}

Note that Eq. (\ref{hj}) can also be written as
\begin{eqnarray}
\left(H - \frac{\phi}{\phi_c^2 - \phi^2}\frac{1}{N} \dot{\phi}
\right)^2 - \frac{\phi_c^2 (\phi_s^2 - \phi^2)}{6 \xi \phi_s^2
(\phi_c^2 - \phi^2)^2} \left(\frac{1}{N}\dot{\phi}\right)^2 -
\frac{1}{3\xi (\phi_c^2 - \phi^2)} V =0. \label{hje}
\end{eqnarray}
There is no stable solution for $\phi>\phi_{s}$ in an isotropic
flat spacetime \cite{futamase89}. This can be understood because
each term in Eq. (\ref{hje}) becomes positive definite if
$\phi>\phi_s (> \phi_c)$. Thus the Hamiltonian constraint cannot
be not satisfied for $\phi>\phi_s$.

\section{Slow-roll approximation \label{sec_slow}}

\begin{figure}[t]
\includegraphics[width=0.55\linewidth,height=0.25\textheight ]{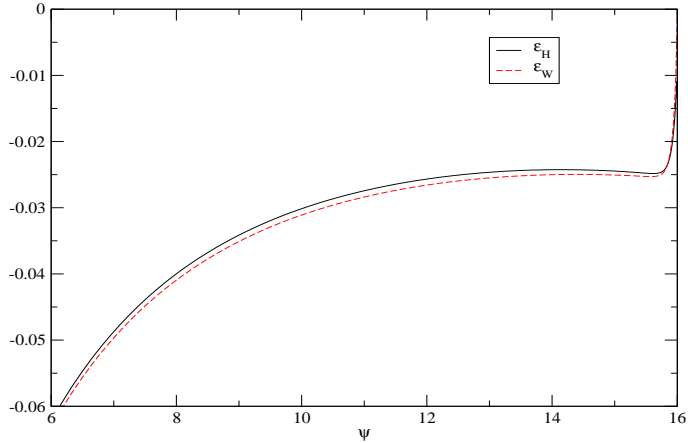}
\caption{Comparison of $\varepsilon_H$ and $\varepsilon_W$ during
an inflation period for $V = m^2\phi^2/2$. Here $\psi \equiv
\sqrt{8\pi G}\phi$ and we set $\psi_0 =16$ and $\xi=10^{-3}$.}
\label{figslow}
\end{figure}

Although the Hamilton-Jacobi method is a powerful tool for
nonlinear evolutions of inhomogeneous spacetimes, it is known to
be difficult to solve exactly the Hamilton-Jacobi equation for a
general potential even in Einstein gravity \cite{salopek90}. In
Ref. \cite{koh05} for generalized gravity, the Hamilton-Jacobi
equation is analytically solved for general potentials using some
approximate methods. A good approximation is the slow-roll
approximation in inflation scenario to get analytical results
which are well fitted to observations. We will try to get an
analytical result of Eq. (\ref{hj}) using the slow-roll
approximation.

The slow-roll conditions during the inflation period are
\begin{eqnarray}
\frac{1}{2}\dot{\phi}^2 \ll V(\phi), \quad
\ddot{\phi} \ll 3 H \dot{\phi}.
\end{eqnarray}
It turns out convenient to define slow-roll parameters from the
above slow-roll conditions
\begin{eqnarray}
\varepsilon_H &\equiv& \frac{\dot{H}}{NH^2} = \frac{\dot{\phi}}{N}
\frac{H_{,\phi}}{H^2}, \quad
\eta_H \equiv \frac{1}{N}\left(\frac{\dot{\phi}}{N}\right)^{\cdot}
\frac{N}{H\dot{\phi}}, \\
\varepsilon_W &\equiv& \frac{\dot{W}}{NW^2} = \frac{\dot{\phi}}{N}
\frac{W_{,\phi}}{W^2}, \quad
\eta_W \equiv \frac{1}{N}\left(\frac{\dot{\phi}}{N}\right)^{\cdot}
\frac{N}{W\dot{\phi}}.
\end{eqnarray}
Here $\varepsilon_H$ and $\eta_H$ are usual definitions of
slow-roll parameters, and $\varepsilon_H$ and $\varepsilon_W$ are
related to each other
\begin{eqnarray}
\varepsilon_H = \frac{\varepsilon_W}{\sqrt{1-8\pi G\xi\phi^2}}
+\frac{2\xi \phi(1-8\pi G \xi \phi^2)}{1-8\pi G\xi\phi^2(1-6\xi)}
\frac{W_{,\phi}}{W}.
\label{slow_rel}
\end{eqnarray}
The occurrence of inflation requires $\varepsilon_H, \eta_H \ll
1$. In Fig. \ref{figslow}, $\varepsilon_H$ and $\varepsilon_W$ are
plotted for $V = m^2 \phi^2 /2$ during the inflation period.
$\varepsilon_H$ and $\varepsilon_W$ are slightly different from
each other, but they are both useful in determining the end of
inflation, $\varepsilon_H \simeq \varepsilon_W =1$.

From Eq. (\ref{hj}), unless $|\xi|$ is much larger than $1$, the
slow-roll condition is equivalent to neglecting the kinetic energy
term in comparison to the potential energy. Then the
Hamilton-Jacobi equation and the evolution equation for $\alpha$
become approximately
\begin{eqnarray}
W^2 &\simeq& \frac{8\pi G}{3(1-8\pi G\xi \phi^2)^2} V
\label{slowhj}, \\
\frac{1}{N}\dot{\alpha} &\simeq& \sqrt{1-8\pi G\xi\phi^2}W.
\label{slowalpha}
\end{eqnarray}

Another useful quantity to describe a sufficient inflation is the
number of $e$-folds, $\mathcal{N}$, which is defined by
\begin{eqnarray}
\mathcal{N}
%= \ln\frac{a}{a_f}
=\int^{\alpha_e}_{\alpha_0}d\alpha = \alpha(\phi_e) - \alpha
(\phi_0)
%=\int^t_{t_f}NH dt
= \int^{\phi_e}_{\phi_0}\frac{H}{\dot{\phi}/N}d\phi,
\label{nefold}
\end{eqnarray}
where subscripts ``$e$'' denotes the end of inflation. To resolve
the cosmological problems such as the flatness, horizon, and
homogeneity problems, a bound $\mathcal{N} \gtrsim 60$ is
required. With the help of Eqs. (\ref{slowalpha}) and
(\ref{field_eq}), $\mathcal{N}$ can be written as
\begin{eqnarray}
\mathcal{N} = -4\pi G\int^{\phi_e}_{\phi} \frac{1-8\pi
G\xi\phi^2(1-6\xi)}{(1-8\pi G\xi\phi^2)^2}
\frac{W}{W_{,\phi}}d\phi.
\end{eqnarray}

\begin{figure}[t]
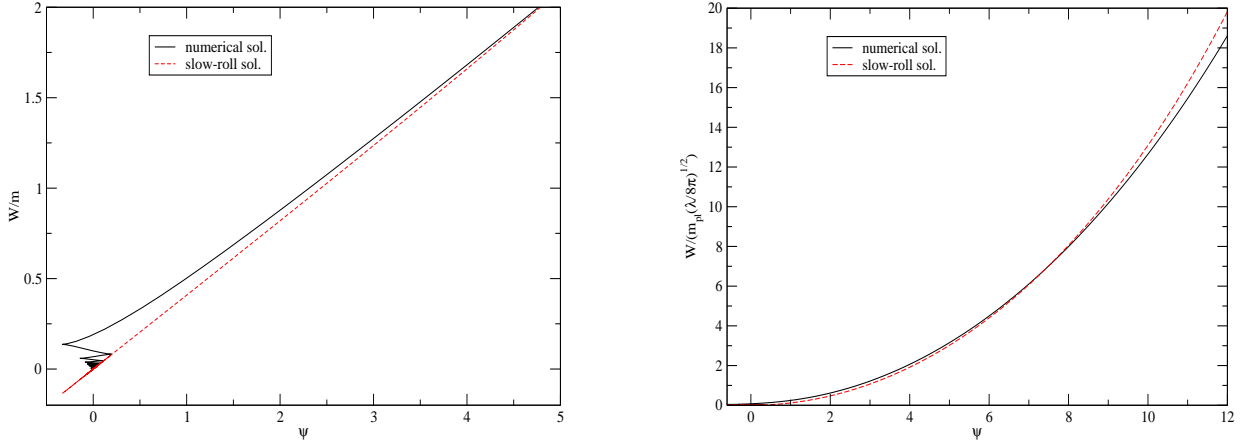

%\subfloat[]
{\includegraphics[width=0.45\linewidth,height=0.25\textheight ]
{slowv2.eps}} \hfill
%\subfloat[]
{\includegraphics[width=0.45\linewidth,height=0.25\textheight
]{slowv4.eps} } \caption{Comparison of slow-roll
solutions with numerical solutions for $V=m^2 \phi^2/2$ (left
panel) and for $V=\lambda \phi^4/4!$ (right panel).}
\label{slowfig}
\end{figure}

We will solve the Hamilton-Jacobi equation for various type
potentials using the slow-roll approximation and find initial
conditions for the scalar field to result in a sufficiently enough
inflation. According to Ref. \cite{dodelson97}, inflation
potentials in a single field model are classified into three types
depending on the shape of potentials and the initial condition of
the scalar field - a large field, small field and hybrid field
model. In the large field model, for example in the chaotic
inflation, the scalar field that is initially greater than
$m_{pl}$ rolls slowly down to the minimum and then oscillates
around the minimum value. And the potential has a positive
curvature at the minimum. Contrary to the large field model, the
scalar field in the small field model such as the new inflation
stays initially near a false vacuum and then evolves toward the
true vacuum. So it is possible to have an inflation even when the
field is much smaller than $m_{pl}$ and the potential at the
minimum has a negative curvature. And finally the scalar field in
the hybrid inflation evolves toward a minimum of the potential
which has a nonzero vacuum energy.

\subsection{A large field model, $V(\phi)\propto \phi^p$}

\begin{figure}[t]
%\subfloat[]
{\includegraphics[width=0.45\linewidth,height=0.35\textheight ]
{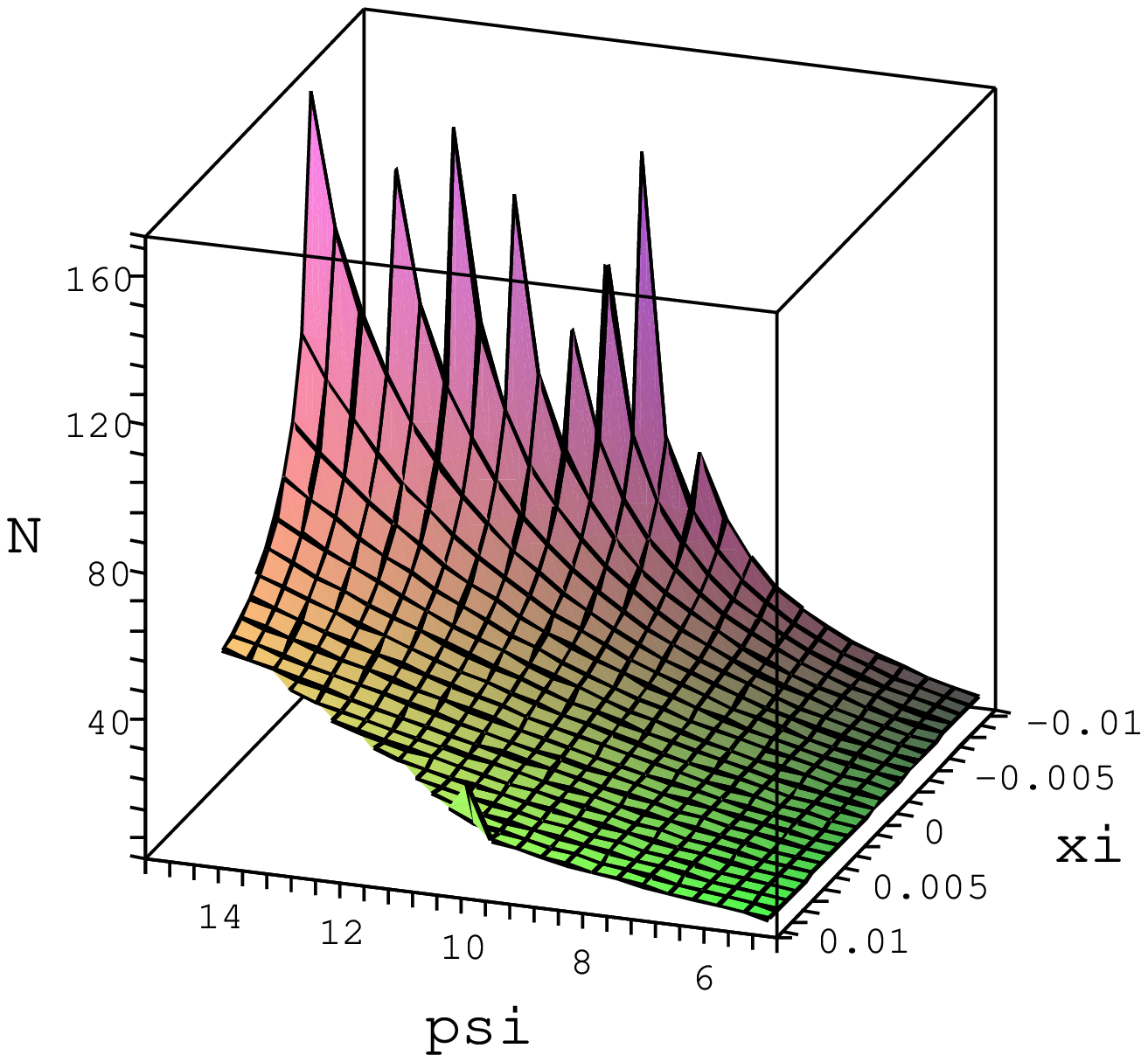}}\hfill 
%\subfloat[]
{\includegraphics[width=0.45\linewidth,height=0.35\textheight ]
{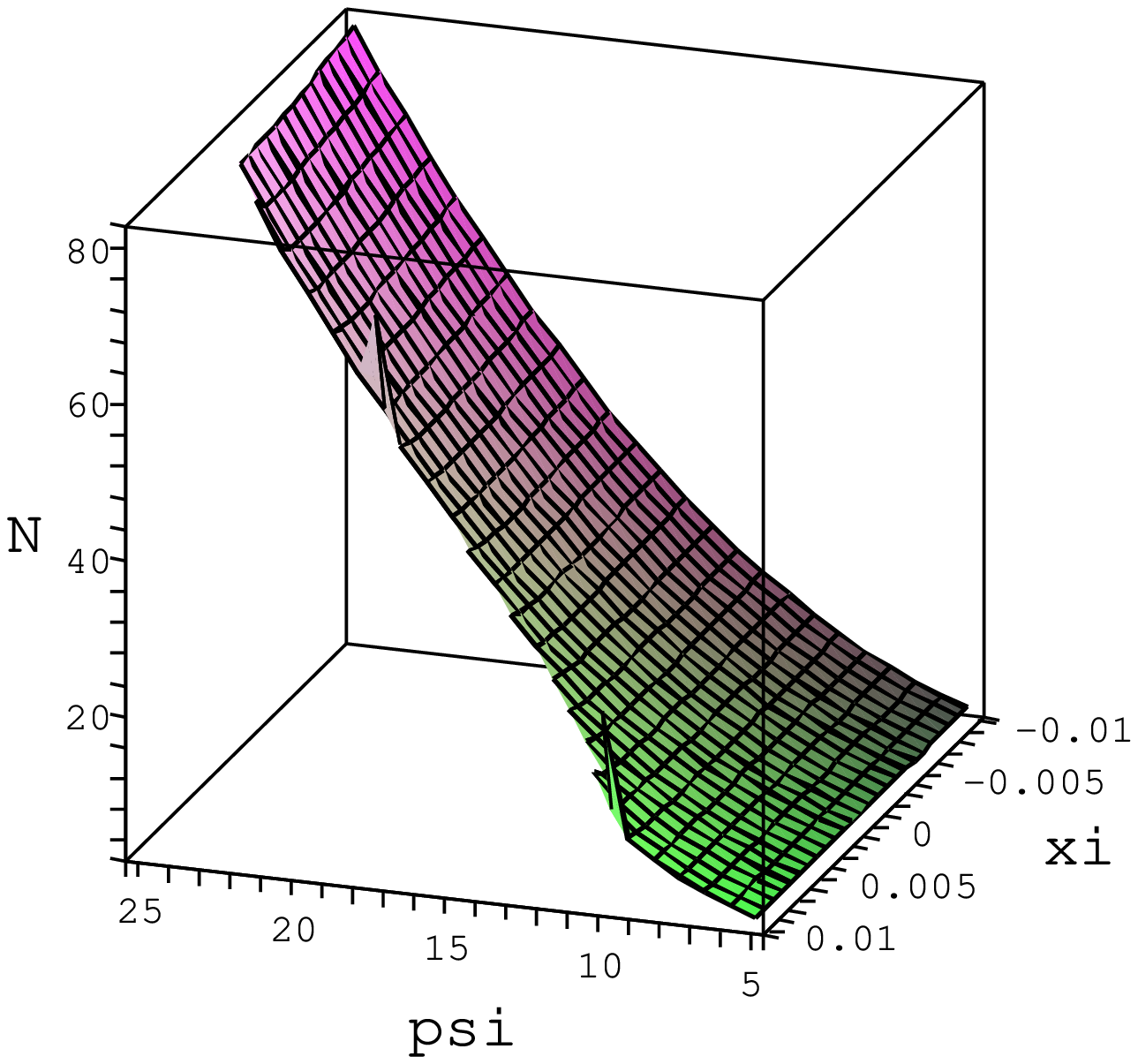}} \caption{The number of
e-folds for ranges of $\xi$ and $\psi_0$ for $V=m^2 \phi^2/2$
(left panel) and for $V=\lambda \phi^4/4!$ (right panel).}
\label{nefoldfig_large}
\end{figure}

An initial condition $\phi_{0} >m_{pl}$ is necessary for a
sufficient inflation to occur in the large field model. The
chaotic inflation belongs to this type. We consider $V=m^2
\phi^2/2$ and $V=\lambda \phi^4/4!$ for the large field model. In
Fig. \ref{slowfig}, 
we compare the exact numerical
solution and the slow-roll approximate solution of the
Hamilton-Jacobi equation for $V=m^2 \phi^2/2$ and $V=\lambda
\phi^4/4!$. For $V=m^2 \phi^2/2$, Eqs. (\ref{slowhj}) and
(\ref{slowalpha}) together with Eq. (\ref{field_eq}) become
\begin{eqnarray}
W &=& \sqrt{\frac{4\pi G}{3}}\frac{m\phi}{1-8\pi G\xi \phi^2}, \\
\frac{d\alpha}{d\phi} &=& -4\pi G\frac{\phi(1-8\pi G\xi\phi^2(1-6\xi))}
{1-64\pi^2 G^2 \xi^2 \phi^4}.
\label{alpha_v2}
\end{eqnarray}

Integrating Eq. (\ref{alpha_v2}) over $\phi$ between $\phi_0$ and
$\phi_{e}$, we can obtain the number of $e$-folds
\begin{eqnarray}
\mathcal{N} = -\frac{1}{4\xi} ({\rm arctanh}(\phi_{e}^2/\phi_c^2)
- {\rm arctanh}(\phi_0^2/\phi_c^2)) -\frac{1-6\xi}{8\xi}
\ln\frac{1-(\phi_{e}^2/\phi_c)^2}{1-(\phi_0^2/\phi_c^2)^2}.
\label{nefold_v2}
\end{eqnarray}
After passing through the slow-roll regime, $\phi$ starts to
oscillate around $\phi_{e} = 0$. The number of $e$-folds are
plotted in Fig. \ref{nefoldfig_large} (left) against $\psi_0 \equiv
\sqrt{8\pi}\phi_0/m_{pl}$ and $\xi$. In addition to the constraint
from  $G_{eff}>0$ for $\xi>0$, $\phi^2/\phi_c^2$ is further
constrained by the form of ${\rm arctanh}$ function, independently
of sign of $\xi$,
\begin{eqnarray}
-1<\phi^2 / \phi_c^2 <1.
\end{eqnarray}

\begin{figure}[t]
\includegraphics[width=0.55\linewidth,height=0.25\textheight ]
{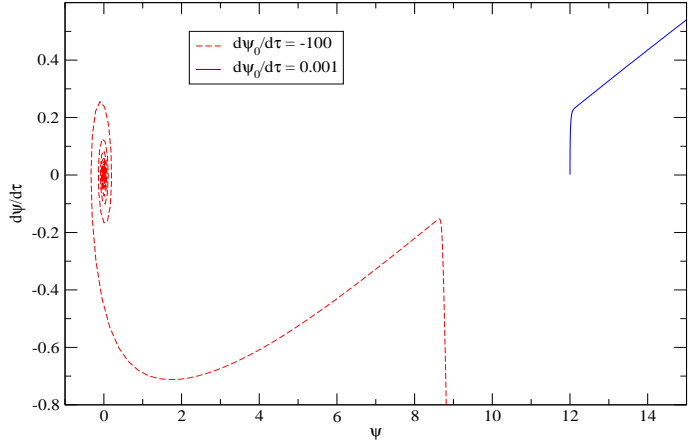} \caption{If $\xi$ is negative and $\phi_0
>\phi_m$, the orbit of the scalar field depends on the initial
value of $\dot{\phi}$ for $V=m^2 \phi^2/2$. For $\xi=-10^{-2}$,
$\psi_0 \equiv \sqrt{8\pi G}\phi_0 =12$, and $d\psi/d\tau =
-100$, $\phi$ crosses over the peak of potential and evolves
toward the origin. The origin is an attractor.}
\label{v2_negative}
\end{figure}

For $\xi <0$, the scalar field shows a different behavior
depending upon the initial value $\phi_0$. If $\phi_0 < \phi_m
\equiv m_{pl}/\sqrt{8\pi |\xi|} = |\phi_c|$, the field $\phi$
evolves toward the origin. But if $\phi_0 > \phi_m$, $\phi$ runs
away to infinity and never reaches the origin. This constraint for
$\xi<0$ is understood in the Einstein frame
\cite{futamase89,tsujikawa00}. The potential in the Einstein frame
can be obtained through a conformal transformation as
\begin{eqnarray}
\hat{V}(\phi) = \frac{V(\phi)}{(1 - \phi^2/\phi_c^2)^2}.
\label{ein_pot}
\end{eqnarray}
The potential $\hat{V}$ has a local maximum at $\phi_{m}$ for
$V=m^2 \phi^2/2$ when $\xi$ is negative. If $\phi_0>\phi_m$, the
field diverges to larger values as shown in Fig. \ref{v2_negative}
instead of rolling down to the origin. But if the initial value of
$\dot{\phi}$ is very large and negative, the field crosses over
the potential barrier and then reaches the origin. Even though it
is possible for $\phi$ field to reach the origin when $\phi_0
>\phi_m$ for $-10^{-3} < \xi < 0$, the
number of $e$-folds is not greater than $60$ because the slow-roll
starts at $\phi \simeq \phi_m$. Thus, the initial condition
$\phi_0$ should be less than $m_{pl}/\sqrt{8\pi |\xi|}$,
independently of the sign of $\xi$. And a constraint $|\xi| \leq
10^{-3}$ is required for a sufficient inflation to occur. In this
case, the first term in Eq. (\ref{nefold_v2}) dominates
\begin{eqnarray}
\mathcal{N} \simeq \frac{1}{4\xi} {\rm
arctanh}(\phi_0^2/\phi_c^2).
\end{eqnarray}
For $\mathcal{N} \gtrsim 60$, $\psi_0 \equiv
\sqrt{8\pi}\phi/m_{pl} \gtrsim 16.5$ when $\xi=10^{-3}$.

\begin{figure}[t]
%\subfloat[]
{\includegraphics[width=0.45\linewidth,height=0.35\textheight ]
{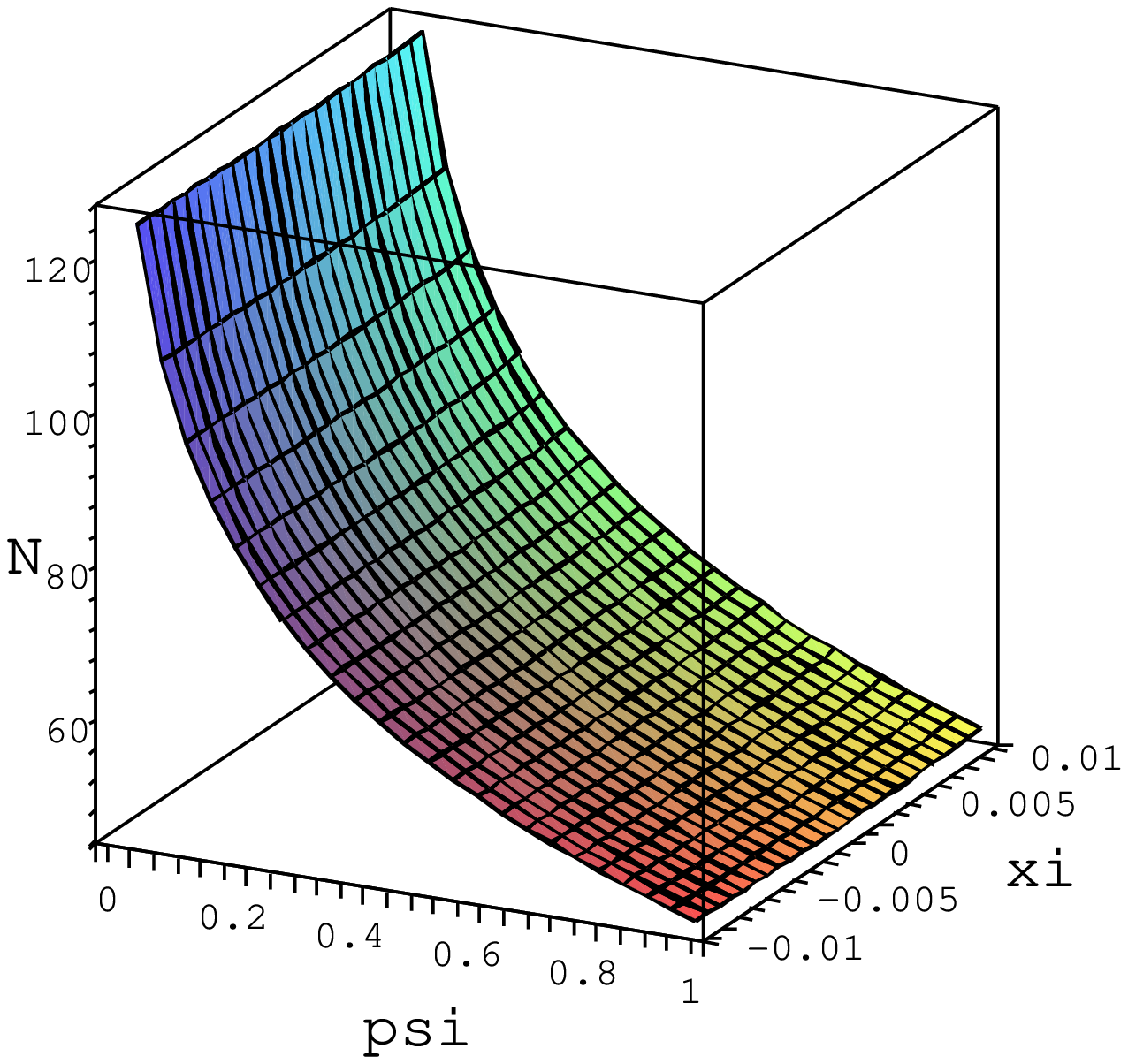}}\hfill 
%\subfloat[]
{\includegraphics[width=0.45\linewidth,height=0.35\textheight ]
{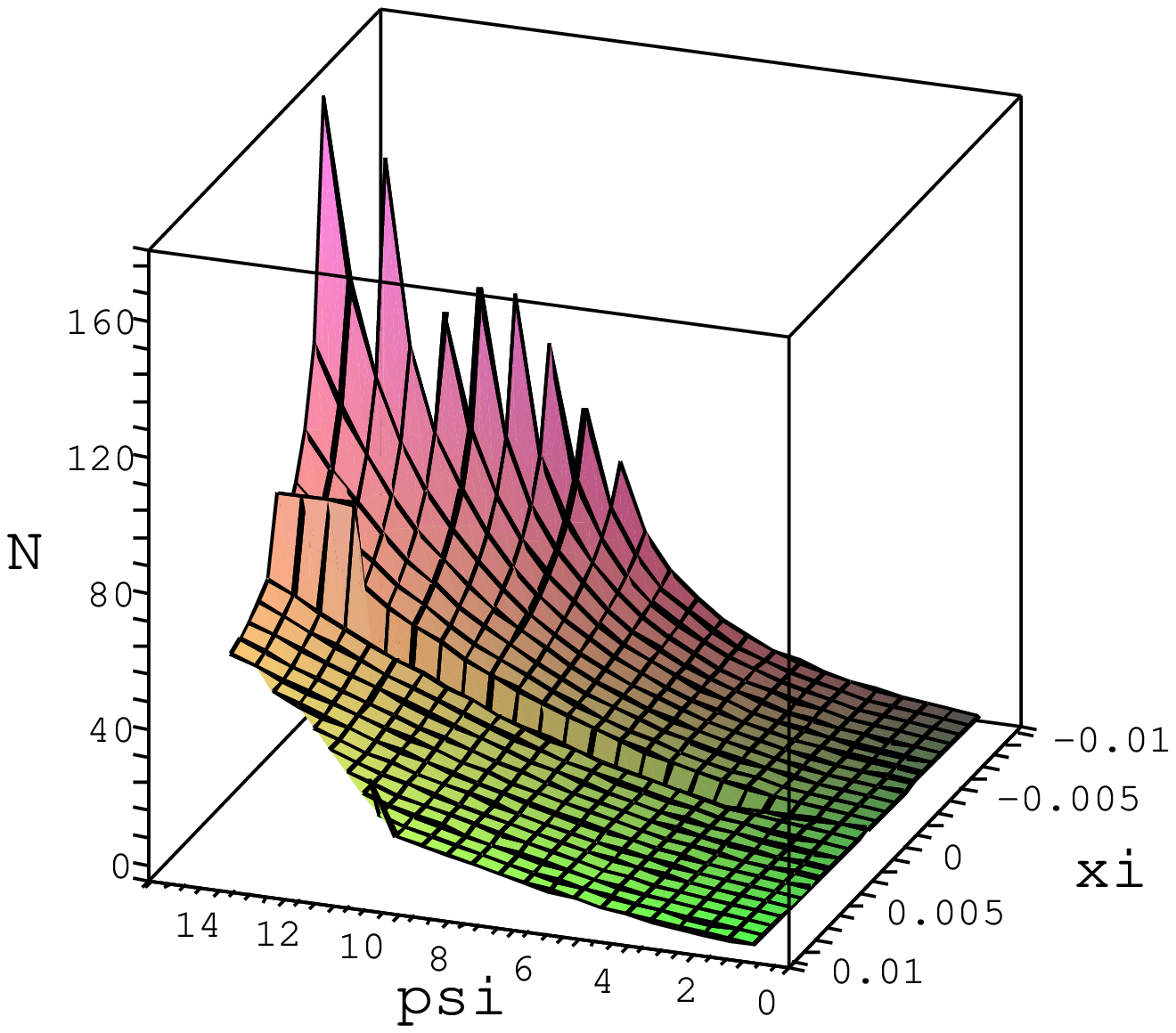}} 
\caption{The number
of e-folds for ranges of $\xi$ and initial values $\psi_0$ in the
small field model (left panel) and in the hybrid model (right
panel).}
\label{nefoldfig}
\end{figure}

Next, we consider $V=\lambda \phi^4/4!$ and then Eqs.
(\ref{slowhj}) and (\ref{slowalpha}) become
\begin{eqnarray}
W &=& \frac{\sqrt{\lambda \pi G}}{3}\frac{\phi^2}{1-8\pi G\xi\phi^2}, \\
\frac{d\alpha}{d\phi} &=& -2\pi G
\frac{\phi(1-8\pi G\xi\phi^2(1-6\xi))}{1-8\pi G\xi \phi^2}.
\label{alpha_v4}
\end{eqnarray}
By simply integrating Eq. (\ref{alpha_v4}) we can get the number
of $e$-folds
\begin{eqnarray}
\mathcal{N} = - \frac{1-6\xi}{8 \xi \phi_c^2} (\phi_{e}^2-\phi_0^2)
+\frac{3}{4}\ln \frac{1 - \phi_{e}^2/\phi_c^2}{1 -
\phi_0^2/\phi_c^2}. \label{nefold_v4}
\end{eqnarray}
We plot the number of $e$-folds in Fig. \ref{nefoldfig_large} (right). For
$\xi>0$, $\phi$ should be less than $\phi_c$ to satisfy the
condition $G_{eff}>0$. A constraint $|\xi|<10^{-3}$ is also
required in order to lead to a sufficient number of $e$-folds as
for the case of $V = m^2\phi^ / 2$. A local maximum of the
potential in the Einstein frame does not exist in this potential.
Thus, there is no constraint on $\phi_0$ even when $\xi$ is
negative. For $|\xi|<10^{-3}$, the first term in Eq.
(\ref{nefold_v4}) dominates the second logarithmic term due to the
factor of $(1-6\xi)/8\xi$. So for $|\xi| \leq 10^{-3}$ and
$\phi_{e} \simeq 0$, we have initial conditions for
$\mathcal{N}\geq 60$
\begin{eqnarray}
\phi_0 \geq \sqrt{\frac{60}{8\pi}}\sqrt{\frac{8}{1-6\xi}} m_{pl}.
\label{bound_v4}
\end{eqnarray}

\subsection{A small field model, $V(\phi) \propto 1-(\phi/\mu)^p$}

\begin{figure}[t]
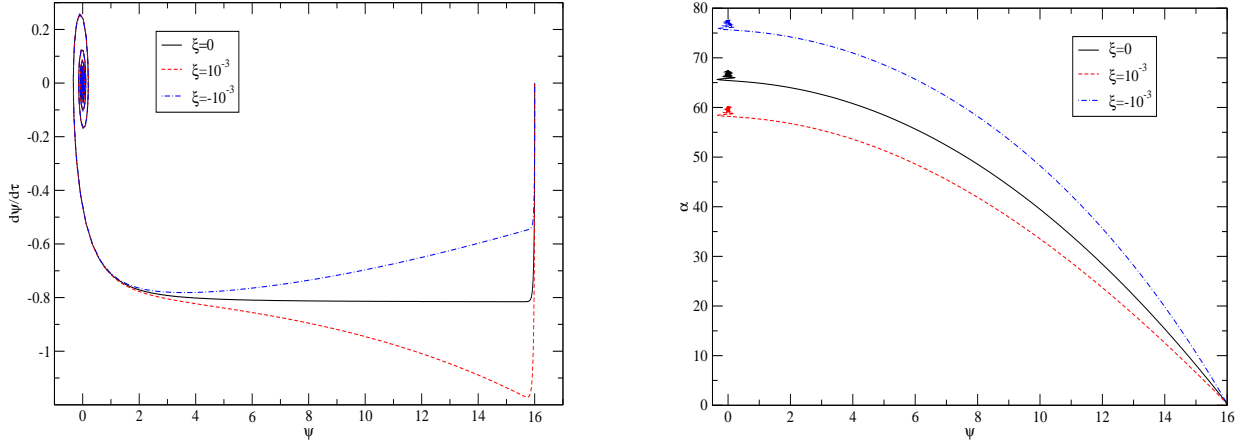

%\subfloat[]
{\includegraphics[width=0.45\linewidth,height=0.25\textheight ]
{v2_phase.eps} } \hfill 
%\subfloat[]
{\includegraphics[width=0.45\linewidth,height=0.25\textheight ]
{v2_alpha.eps} } \caption{Phase diagram of
$\psi \equiv \sqrt{8\pi G}\phi$ and $d\psi/d\tau$ (left
panel) and $\alpha$ against $\psi$ (right panel) for $V(\phi)= m^2
\phi^2/2$. An initial value $\psi_0 = 16$ is taken.}
\label{v2fig}
\end{figure}

Let us consider a Ginzburg-Landau potential
\begin{eqnarray}
V(\phi) = \frac{\chi}{8}(\phi^2-\mu^2)^2.
\label{small_pot}
\end{eqnarray}
If $|\phi| \gg \mu$, this potential can be regarded as a large
field model with a term $\phi^4$ \cite{fakir90}. Therefore we
assume that $|\phi| \ll \mu$. In the vicinity of the origin, this
potential is approximated as
\begin{eqnarray}
V(\phi) =\Lambda^4 \left( 1 - \left(\frac{\phi}{\mu}\right)^2\right).
\end{eqnarray}
where $\Lambda = \mu (\chi/8)^{1/4}$. The field $\phi$ is
initially located around the origin and then slowly rolls down to
the  true vacuum at $\mu =\langle \phi \rangle$. With the
slow-roll condition, the Hamilton-Jacobi equation and the
evolution equation for $\alpha$ become
\begin{eqnarray}
W &=& \sqrt{\frac{\pi\chi G}{3}}\frac{\phi^2-\mu^2}{1-8\pi G\xi\phi^2}, \\
\frac{d\alpha}{d\phi} &=& -\frac{2\pi G}{1-8\pi G\xi\mu^2}
\frac{(\phi^2-\mu^2)(1-8\pi G\xi\phi^2 (1-6\xi))}
{\phi(1-8\pi G\xi \phi^2)}.
\label{slowalpha_small}
\end{eqnarray}
Then, the number of $e$-folds is
\begin{eqnarray}
\mathcal{N} = \frac{1}{4(\phi_c^2-\mu^2)}\Biggl[
-\frac{1-6\xi}{2\xi}(\phi_{e}^2-\phi_0^2)
+\frac{\mu^2}{\xi}\ln\frac{\phi_{e}}{\phi_0}
+3(\phi_c^2- \mu^2)
\ln\frac{1-\phi_{e}^2/\phi_c^2} {1-\phi_0^2/\phi_c^2} \Biggr].
\label{nefold_small}
\end{eqnarray}
We assume that inflation ends when $\phi_{e} \simeq \mu$. In Eq.
(\ref{nefold_small}), if $1-\mu^2/\phi_c^2 <0$,
$\phi_0^2/\phi_c^2$ should be larger than $1$. However, this
violates the condition $G_{eff}>0$ in Eq. (\ref{eff_grav})  for
$\xi>0$. So we restrict
 to $\mu^2/\phi_c^2 <1$ for $\xi>0$. If $\phi \ll m_{pl}$,
the second term in Eq. (\ref{nefold_small}) dominates in the
number of e-foldings
\begin{eqnarray}
\mathcal{N} \simeq \frac{\mu^2}{4\xi(\phi_c^2-\mu^2)}\ln
\frac{\mu}{\phi_0}.
\end{eqnarray}
In Fig. \ref{nefoldfig} (left), we plot the number of $e$-foldings
for various ranges of $\xi$ and $\psi\equiv \sqrt{8\pi G}\phi_0$
and $\sqrt{8\pi G}\mu=10$.

Note that the nonminimal coupling might prevent inflationary
solutions in the new inflation \cite{abbott81} because the
coupling term, $\xi R\phi^2/2$, behaves like a mass term in the
scalar field potential and destroys the flatness of the potential
near the origin. However, we will examine a possibility of the
inflationary solutions in the new inflation when $\phi \ll \mu$
and $8\pi G\xi \phi^2 \ll 1$. During the slow-roll of the
inflation, the potential (\ref{small_pot}) can be approximated as
$V \simeq \Lambda^4$. Then Eqs. (\ref{field_eq}) and
(\ref{slowalpha_small}) are approximated by
\begin{eqnarray}
\dot{\phi} &\simeq& \sqrt{\frac{128\pi G\Lambda^4}{3}}\xi \phi, \\
\frac{d\alpha}{d\phi} &\simeq& \frac{2\pi G\mu^2}{1-8\pi G\xi\mu^2}
\frac{1}{\phi}.
\end{eqnarray}
where we have set $N=1$. Thus we obtain the solutions
\begin{eqnarray}
\phi = \phi_i \exp \left[\sqrt{\frac{128\pi G\Lambda^4}{3}}\xi t\right],
\quad \alpha =\alpha_i \sqrt{\frac{8\pi G\Lambda^4}{3}}\frac{8\pi G\xi\mu^2}
{1-8\pi G\xi\mu^2}t,
\end{eqnarray}
where $\phi_i$ and $\alpha_i$ are integration constants. Because
$\alpha$ is a logarithm of the scale factor, this shows that the
universe expands quasi-exponentially when the nonminimal coupling
exists. But here we do not consider the dynamics of the phase
transition in detail during the slow-roll phase, so it requires
scrutiny to conclude about the existence of inflationary solutions
in the new inflationary scenario.

\subsection{A hybrid field model, $V \propto 1+(\phi/\mu)^p$}

\begin{figure}[t]
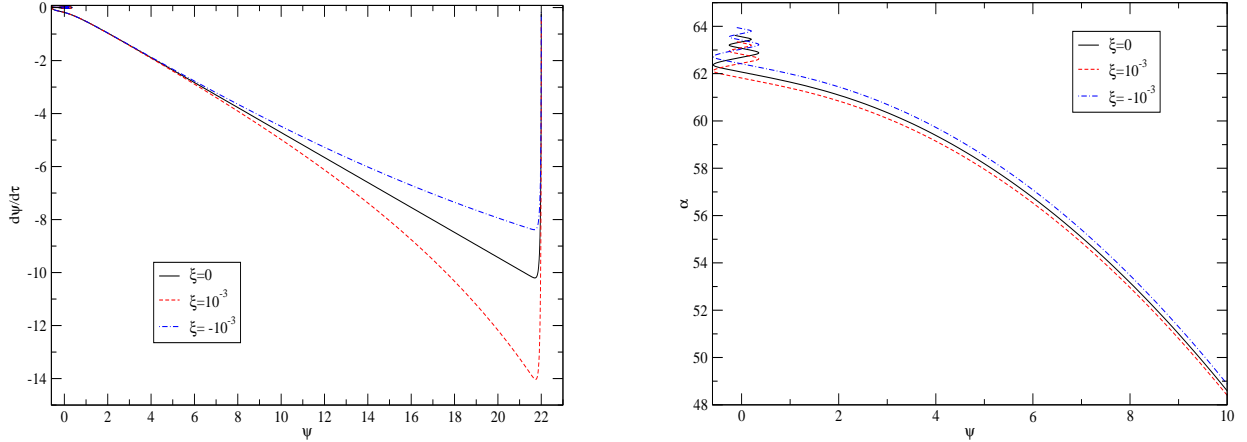

%\subfloat[]
{\includegraphics[width=0.45\linewidth,height=0.25\textheight ]
{v4_phase.eps} }\hfill 
%\subfloat[]
{\includegraphics[width=0.45\linewidth,height=0.25\textheight ]
{v4_alpha.eps} } \caption{Phase diagram of $d
\psi/d \tau$ and $\psi$ (left panel) and $\alpha$ against $\psi$
(right panel) for $V(\phi)= \lambda \phi^4/4!$. The initial value
$\psi_0 =22$ is taken.}
\label{v4fig}
\end{figure}

This type of potential describes the hybrid inflation
\cite{linde91, copeland94}. Contrary to the other inflation
models, hybrid inflation gives a blue power spectrum. The
potential in hybrid inflation may take the form
\begin{eqnarray}
V(\sigma,\phi) = \frac{\lambda}{4}(M^2-\sigma^2)^2
+\frac{m^2}{2}\phi^2 +\frac{g^2}{2}\phi^2 \sigma^2.
\end{eqnarray}
When $\phi^2 > \phi_{f}^2 \equiv \lambda M^2/g$, there is a local
minimum at $\sigma=0$ (false vacuum) on the constant $\phi$
slices. Inflation begins when $\sigma$ is located at the false
vacuum ($\sigma =0$), and then the $\phi$ field rolls down toward
the origin ($\phi=0$). During the inflation period and near
$\sigma = 0$, the potential can be given by
\begin{eqnarray}
V(\phi) = \frac{\lambda}{4}M^4 +\frac{1}{2}m^2 \phi^2
\label{hyb_pot2}.
\end{eqnarray}

\begin{figure}[t]
\includegraphics[width=0.55\linewidth,height=0.25\textheight ]
{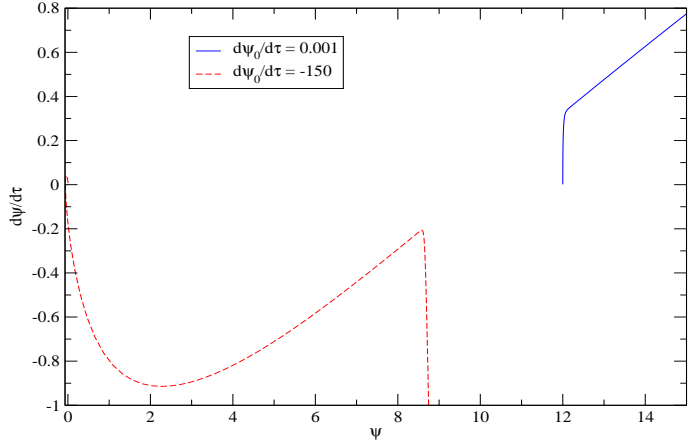} \caption{When $\xi<0$ and $\phi_0 >\phi_m$,
the orbit of the scalar field depends upon the initial value of
$\dot{\phi}$ for $V=\Lambda^4(1+(\phi/\mu)^2)$. For
$\xi=-10^{-2}$, $\psi_0 =12$, and $d\psi/d\tau = -150$, $\phi$
crosses over the peak of potential and evolves toward the origin
as in the case of $V=m^2 \phi^2/2$.} \label{hybrid_negative}
\end{figure}

If the false vacuum energy dominates the potential energy,
inflation ends when $\phi^2<\phi_{f}^2$. The hybrid field model
differs from the large field model in the sense that during
inflation the vacuum energy is nonzero. The potential,
(\ref{hyb_pot2}), can be rewritten as
\begin{eqnarray}
V(\phi) = \Lambda^4 \left(1+\left(\frac{\phi}{\mu}\right)^2\right),
\end{eqnarray}
where $\Lambda^4 = \lambda M^4/4$ and $\mu^2 = \lambda M^4/2m^2$.
The Hamilton-Jacobi equation and the evolution equation for
$\alpha$ become
\begin{eqnarray}
W &=& \sqrt{\frac{8\pi G\Lambda^4}{3\mu^2}}
\frac{\sqrt{\mu^2+\phi^2}}{1-8\pi G\xi\phi^2}, \\
\frac{d\alpha}{d\phi} &=& -4\pi G
\frac{(\mu^2+\phi^2)(1-8\pi G\xi\phi^2(1-6\xi))}
{\phi(1-8\pi G\xi\phi^2)(1+16\pi G\xi\mu^2+8\pi G\xi\phi^2)}.
\label{hybrid_alpha}
\end{eqnarray}
By simply integrating Eq. (\ref{hybrid_alpha}), we obtain
\begin{eqnarray}
\mathcal{N} &=& -\frac{1}{8\xi(1+ 2\mu^2/\phi_c^2)}\Biggl[ 4
\frac{\mu^2}{\phi_c^2} \ln\frac{\phi_{e}}{\phi_0}
-6\xi\left(1 + \frac{2 \mu^2}{\phi_c^2}\right)
\ln\frac{1 - \phi_{e}^2/\phi_c^2}{1 - \phi_0^2/\phi_c^2} \nonumber \\
& &+\left(2-6\xi + \frac{2 \mu^2}{\phi_s^2}\right)
\ln \frac{1+ 2\mu^2/\phi_c^2 +
\phi_{e}^2/\phi_c^2}{1+ 2\mu^2/\phi_c^2 +
\phi_0^2/\phi_c^2}\Biggr]. \label{nefold_hybrid}
\end{eqnarray}

We assume that the inflaton field $\phi$ dominates the vacuum
energy but the $\sigma$ field plays no significant role during the
inflation period. This implies that $\mu \ll m_{pl} \ll \phi$
\cite{copeland94}. With this assumption, the third term in Eq.
(\ref{nefold_hybrid}) is dominant compared to the other two terms.
 If we set $\phi_{e}\simeq 0$ and
for $\xi<0$, note that the numerator in the second logarithmic
term is greater than zero, then $8\pi G|\xi|\phi_0^2$ should be
less than $1$. With the potential given in (\ref{hyb_pot2}), this
situation is similar to the large field model with $V=m^2
\phi^2/2$. When $\xi$ is negative, the maximum value of the
potential in the Einstein frame, (\ref{ein_pot}), which could be
obtained through the conformal transformation, is located at
\begin{eqnarray}
\phi = \phi_m \equiv \frac{1-16\pi G |\xi|\mu^2}{8\pi G|\xi|}.
\end{eqnarray}
If $8\pi G|\xi|\mu^2 \ll 1$, then $\phi_m \simeq
m_{pl}/\sqrt{8\pi|\xi|}$. Similarly to the case of $V=m^2
\phi^2/2$, even if the $\phi_0$ is greater than $\phi_m$, the
$\phi$ field crosses over the barrier of the potential and then
evolves toward the origin when it has a very large negative
initial velocity. It is, however, required $\phi_0$ should be less
than $\phi_m$ and $|\xi|\leq 10^{-3}$ as stated in the case of
$V=m^2 \phi^2 /2$. The phase diagram for $\xi<0$ and $\phi_0
>\phi_m$ are plotted in Fig \ref{hybrid_negative}. For $\mu\ll
m_{pl}$, Eq. (\ref{nefold_hybrid}) could be approximated as
\begin{eqnarray}
\mathcal{N} \simeq -\frac{2-6\xi}{8\xi}\ln \frac{1}{1 +
\phi_0^2/\phi_c^2}.
\end{eqnarray}

\section{Numerical Solutions of Hamilton-Jacobi Equation \label{sec_num}}

\begin{figure}[t]
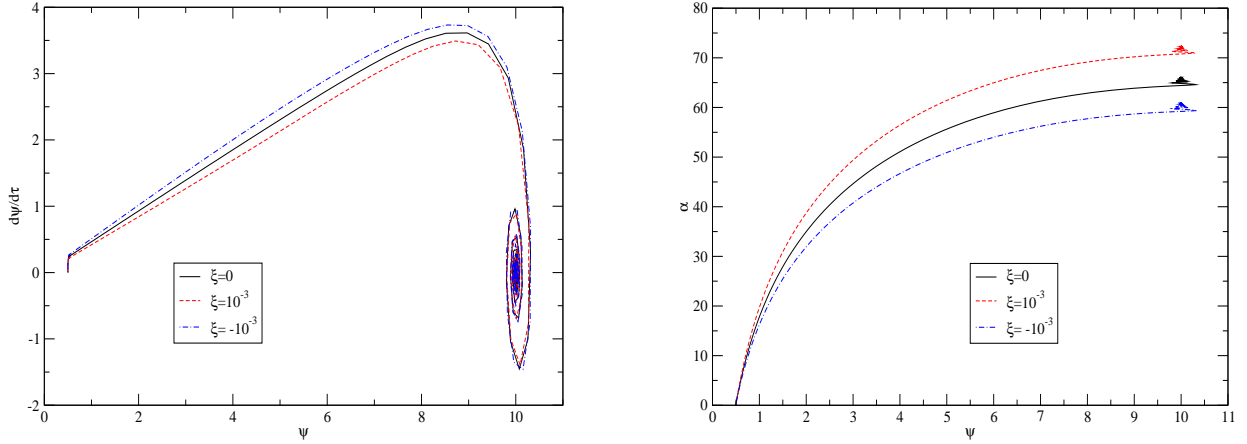

%\subfloat[]
{\includegraphics[width=0.45\linewidth,height=0.25\textheight ]
{small_phase.eps} }\hfill 
%\subfloat[]
{\includegraphics[width=0.45\linewidth,height=0.25\textheight ]
{small_alpha.eps} } \caption{Phase diagram
of $d \psi/d \tau$ and $\psi$ (left panel) and $\alpha$ against
$\psi$ (right panel) for $V(\phi)=\chi(\phi^2-\mu^2)^2/8$. We take
the initial parameters $\psi_0 =0.5$ and $\sqrt{8\pi G}\mu = 10$.}
\label{smallfig}
\end{figure}

In this section, we numerically solve the Hamilton-Jacobi equation
(\ref{hj}) and the evolution equation (\ref{hubble}) for $\alpha$
and Eq. (\ref{field_eq}) for $\phi$ in the nonminimally coupled
scalar field theory. And then we compare this result with that of
Einstein gravity, $\xi=0$ in this paper. $W$ could be interpreted
as a Hubble parameter $H\equiv \dot{\alpha}/N$ for $\xi=0$. We
choose the synchronous gauge ($N=1$) for numerical calculations
and, using Eqs. (\ref{field_eq}) and (\ref{hj}), derive the second
order differential equation for $\phi$

\begin{eqnarray}
\ddot{\phi} &=&
-3\sqrt{1-8\pi G\xi\phi^2}W\dot{\phi}
-\frac{16\pi G\xi\phi}{1-8\pi G\xi\phi^2}\dot{\phi}^2
\left[1-\frac{3\xi}{1-8\pi G\xi\phi^2(1-6\xi)}\right]
\nonumber \\
&-& \frac{1-8\pi G\xi \phi^2}{1-8\pi G\xi\phi^2(1-6\xi)}V_{,\phi}
-\frac{32\pi G\xi\phi}{1-8\pi G\xi\phi^2(1-6\xi)}V.
\end{eqnarray}
The dimensionless quantities to be introduced are
\begin{eqnarray}
\psi = \frac{\sqrt{8\pi}\phi}{m_{pl}},\quad \tau =
\frac{t}{t_{\ast}},
\end{eqnarray}
where $t_{\ast}$ depends on the shape of the potential. We choose
$t_{\ast} = 1/m$ for $V=m^2 \phi^2/2$, $t_{\ast} =
\sqrt{8\pi/\lambda}/m_{pl}$ for $V=\lambda \phi^4/4!$, $t_{\ast} =
\sqrt{8\pi/\chi}/m_{pl}$ for $V=\chi (\phi^2-\mu^2)^2/8$, and
$t_{\ast} = m_{pl}/\sqrt{8\pi}/\Lambda^2$ for
$V=\Lambda^4(1+(\phi/\mu)^2)$.

In Fig. \ref{v2fig}, we plot the phase
diagram and logarithm of the scale factor, $\alpha$, for $\phi$
for $V=m^2 \phi^2/2$ when the initial value is prescribed by
$\psi_0 =16$. From the discussion in the previous section,
$\psi_0^2$ should be smaller than $1/|\xi|$ for $V=m^2 \phi^2/2$.
Therefore, the constraint $|\xi| \lesssim 10^{-3}$ is required for
a sufficient inflation to occur (see Fig. \ref{nefoldfig_large} (left)).
Fig. \ref{v2fig} (left) shows an attractor at
$(\phi,\dot{\phi})=(0,0)$ if the initial condition $\psi_0^2
<1/|\xi|$ is satisfied. The number of $e$-folds in Fig.
\ref{v2fig} (right) decreases for $\xi>0$ compared to the case of
$\xi=0$ but is enhanced for $\xi<0$. As stated in the previous
section, if $\psi$ is greater than $\psi_m = 1/\sqrt{|\xi|}$ for
$\xi<0$, the orbit in the phase diagram may reach the origin or
evolve to larger values of $\psi$ depending on the velocity
$d\psi/d\tau$. This is shown in Fig. \ref{v2_negative}. If
$d\psi/d\tau =10^{-3}$ for $\xi=-10^{-2}$ and for $\psi_0= 12$,
which is larger than $\psi_m = 10$, $\psi$ does not evolve toward
the origin. On the contrary, if $d\psi/d\tau = -10^2$, it crosses
over the peak of the potential and then reaches the attractor.
However, if $|\xi|>10^{-3}$, it does not give a sufficiently
enough inflation.

The phase diagram and number of $e$-folds for $V=\lambda
\phi^4/4!$ are plotted in Fig. \ref{v4fig} 
with $\psi_0=22$. And $W$ against $\phi$ is
plotted in Fig. \ref{slowfig} (right). For $\xi>0$, the initial value of
$\psi$ is constrained by $\psi <1/\sqrt{\xi}$. So to have
$\mathcal{N} \gtrsim 60$, one needs $\xi \lesssim 10^{-3}$. But,
on the contrary to the case of $V=m^2 \phi^2/2$, it is possible to
have inflation for $\xi<0$, irrespective of the magnitude of
$|\xi|$ (see Fig. \ref{nefoldfig_large} (right)). The attractor is shown at
$(\phi,\dot{\phi})=(0,0)$ in Fig. \ref{v4fig} (left). As in the
case of $V=m^2 \phi^2/2$, $\mathcal{N}$ in Fig. \ref{v4fig} (right)
decreases for positive $\xi$ compared to the $\xi=0$ case, whereas
it increases for negative $\xi$. The number of $e$-folds, however,
do not much depend on $\xi$. The initial condition $\psi_0
> 22$ is at least needed for $\xi=10^{-3}$ in Eq.
(\ref{bound_v4}).

\begin{figure}[t]
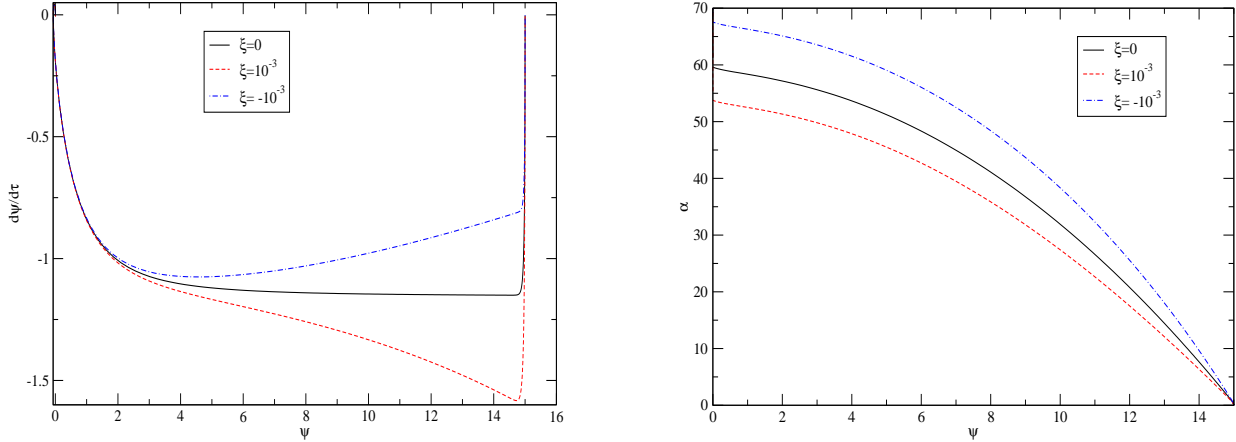

%\subfloat[]
{\includegraphics[width=0.45\linewidth,height=0.25\textheight ]
{hybrid_phase.eps} }\hfill 
%\subfloat[]
{\includegraphics[width=0.45\linewidth,height=0.25\textheight ]
{hybrid_alpha.eps} } \caption{Phase diagram
of $d \psi/d \tau$ and $\psi$ (left panel) and $\alpha$ against
$\psi$ (right panel) for $V(\phi)=\Lambda^4(1+(\phi/\mu)^2)$. We
take the initial parameters as $\psi_0 =15$ and $\sqrt{8\pi
G}\mu=1$.}
\label{hybridfig}
\end{figure}

In Fig. \ref{smallfig}, the phase
space diagram and number of $e$-folds are plotted for the small
field model with the potential (\ref{small_pot}). Contrary to the
large field model, the attractor in phase diagram is located at
$(\phi,\dot{\phi})=(\mu,0)$, where $\mu =\langle \phi\rangle$. We
have chosen $\psi_0 = 0.5$ and $\sqrt{8\pi G}\mu = 10$.
Inflationary solutions are shown in Fig. \ref{smallfig} (right),
which can be seen from $\alpha \propto \ln \phi$ implying a
quasi-exponential expansion of the scale factor. The field is
initially located at the origin ($\psi_0 \ll 1$) and then evolves
to the minimum of the potential, $\mu$. When the field arrives at
the minimum, the inflation ends and the field starts to oscillate
around the minimum. As long as $\psi_0 \ll 1$, there is no
constraint on the $\psi_0$ (see Fig. \ref{nefoldfig} (left)) but
for $\xi>0$, $\mu$ must be less than $1/\sqrt{8\pi G \xi}$. The
number of $e$-folds, $\mathcal{N}$, increases for $\xi>0$ relative
to the $\xi=0$ case, which differs from the case of the large
field model, but it decreases for $\xi<0$.

We plot the phase diagram and number of $e$-folds in Fig.
\ref{hybridfig} for the hybrid
inflation potential. We put the parameters $\psi_0 = 15$ and
$\sqrt{8\pi G}\mu = 1$. Starting from $\psi_0$, the field rolls
down toward the false vacuum, $V = \Lambda^4$. Similarly to the
large field model with $V=m^2 \phi^2/2$, the hybrid inflation also
gives the constraint on $\psi_0$ as $\psi_0^2 <1/|\xi|$. This
constraint comes from the condition $G_{eff}>0$ for $\xi>0$ and
the potential barrier of hybrid inflation in the Einstein frame
which prevents the field from reaching the origin for $\xi<0$. In
Fig. \ref{nefoldfig} (right), $\mathcal{N}$ is plotted for possible
$\psi_0$ and $\xi$. It shows that a constraint of $|\xi| \leq
10^{-3}$ is needed for $\mathcal{N} \gtrsim 60$. Similar arguments
in the case of $V=m^2 \phi^2/2$ are possible for $\xi<0$ when the
field initially is located in the region larger than the peak of
the potential, $\psi_m = 1/|\xi|$. If the initial velocity
($d\psi/d\tau$) of the field is much small, the field evolves to
larger values of $\psi$ and does never arrive at the origin. But
if the field has a large and negative initial velocity, then the
field crosses over the peak of the potential and then evolves to
the origin. We plot the phase diagram for $\xi<0$ when the $\psi
>\psi_m =10$ in Fig. \ref{hybrid_negative}. We compare
$d\psi/d\tau=0.001$ with $-150$ for $\xi =-10^{-2}$.

\section{summary \label{sec_summary}}

Using the Hamilton-Jacobi theory we have studied nonlinear
evolutions of inhomogeneous spacetimes when scalar fields are
nonminimally coupled with the spacetime curvature $R$. And we have
found analytic and numerical solutions for several classes of
inflation potentials during the inflation period. The inflation
potentials thus studied could be categorized into three different
types - a large field, small field, and hybrid field model,
depending on the potential shape and the initial condition of
scalar fields. Though it is believed that in the nonminimally
coupled scalar theory, the new inflation, a small field model, may
not have an inflationary solution, we have shown through an
approximation and numerics that the new inflation potential can
provide a quasi-exponential inflationary solution. Careful
scrutiny is in order before reaching a definite conclusion for the
existence of inflationary solutions. For the hybrid inflation
potential, we have assumed that the inflaton field dominates the
potential. However, if the potential is dominated by the vacuum
energy, which is the limit of the small field model, the other
scalar field, $\sigma$, plays a significant role in terminating
the inflation. It would be interesting to consider the
inflationary solutions for multi-field potentials.

In the case of $\xi>0$, the coupling constant is restricted to a
range of $\xi<10^{-3}$ to have the number of $e$-folds greater
than $60$ whose value is needed to resolve the well known
cosmological puzzles. On the other hand, in the case of $\xi<0$,
$V = \lambda \phi^4/4!$ in the large field or small field model
does not lead to any constraint on $\xi$ to satisfy the
observational bound on $\mathcal{N}$, but $V = m^2 \phi^2/2$ in
the large field model or hybrid field model with $\mu \ll \phi$,
the scaled field $\psi_0 \equiv \sqrt{8\pi}\phi_0/m_{pl}$ should
be smaller than $1/|\xi|$ and $|\xi| <10^{-3}$. If these
potentials in the original frame called Jordan frame are
transformed into those in the Einstein frame by a conformal
transformation, the local maxima of the potentials are located at
$\phi_m =1/\sqrt{8\pi G |\xi|}$ for $V=m^2\phi^/2$ and $\phi_m =
(1-16\pi G|\xi|\mu^2)/8\pi G |\xi|$ for the hybrid field model. If
the field $\phi$ is greater than $\phi_m$, it goes away to
infinity and never reaches the origin. However, if the scalar
field has a very large negative initial velocity, it can cross
over the potential barrier and evolve toward the origin. But the
number of $e$-folds being greater than $60$ restricts to the
nonminimal coupling constant to $|\xi|<10^{-3}$.

It would be more helpful to understand analytically or numerically
the behaviors of nonlinear evolutions of inhomogeneous spacetimes
in the nonminimally coupled scalar field theory beyond the zeroth
order Hamilton-Jacobi equation, which will be addressed in a
future publication.

\acknowledgments This work was supported by Korea Astronomy
Observatory (KAO). The work of SPK was also supported in part by
KOSEF under R01-2005-000-10404-0.


\begin{thebibliography}{999}

\bibitem{maldacena03} J. Maldacena, J. High Energy Phys. {\bf 05}, 013 (2003);
N. Bartolo, E. Komatsu, S. Matarrese, and A. Riotto, Phys. Rep. {\bf 402},
103 (2004).
\bibitem{salopek90} D. S. Salopek and J. R. Bond, Phys. Rev. D {\bf 42},
3936 (1990).
\bibitem{koh05} S. Koh, S. P. Kim, and D. J. Song, astro-ph/0501401

\bibitem{soda95} J. Soda, H. Ishihara, and O. Iguchi, Prog. Theor. Phys.
{\bf 94}, 781 (1995).
\bibitem{lidsey97} J. E. Lidsey {\it et al.}, Rev. Mod. Phys. {\bf 69},
 373 (1997).
\bibitem{dodelson97} S. Dodelson, W. H. Kinney, and E. W. Kolb, Phys. Rev. D
{\bf 56}, 3207 (1997); W. H. Kinney, {\it ibid.} {\bf 58}, 123506 (1998).
\bibitem{linde83} A. D. Linde, Phys. Lett. {\bf B 129}, 177 (1983).
\bibitem{linde82} A. D. Linde, Phys. Lett. {\bf B 108}, 389 (1982);
A. Albrecht and P. J. Steinhardt, Phys. Rev. Lett. {\bf 48}, 1220 (1982).
\bibitem{linde91} A. Linde, Phys. Lett., B {\bf 259},38 (1991); A. D. Linde,
Phys. Rev. D {\bf 49}, 748 (1994).
\bibitem{copeland94} E. J. Copeland, A. R. Liddle, D. H. Lyth,
E. D. Stewart, and D. Wands, Phys. Rev. D {\bf 49}, 6410 (1994).
\bibitem{futamase89} T. Futamase and K. Maeda, Phys. Rev. D {\bf 39},
399 (1989).
\bibitem{fakir90} R. Fakir and W. G. Unruh, Phys. Rev. D {\bf 41}, 1783 (1990).
\bibitem{tsujikawa00} S. Tsujikawa, Phys. Rev. D {\bf 62}, 043512 (2000);
S. Tsujikawa and B. Gumjudpai, Phys. Rev. D {\bf 69},
123523 (2004).
\bibitem{abbott81} L. F. Abbott, Nucl. Phys. {\bf B 185}, 233 (1981);
V. Faraoni, Phys. Rev. D {\bf 53}, 6813 (1996).

\end{thebibliography}
\end{document}